\documentclass[12pt]{article}
\usepackage{amssymb,amsmath,eucal,color}
\usepackage[utf8]{inputenc}

\input cyracc.def
\font\tencyr=wncyr10
\def\cyr{\tencyr\cyracc}

\renewcommand{\a}{\alpha}
\renewcommand{\b}{\beta}

\renewcommand{\d}{\delta}
\newcommand{\e}{\epsilon}
\newcommand{\f}{\varphi}
\newcommand{\g}{\gamma}
\newcommand{\h}{\eta}

\renewcommand{\k}{\kappa}
\renewcommand{\l}{\lambda}
\renewcommand{\o}{\omega}

\renewcommand{\r}{\rho}
\newcommand{\s}{\sigma}

\newcommand{\F}{\Phi}

\newcommand{\La}{\Lambda}
\renewcommand{\O}{\Omega}

\newcommand{\bC}{\mathbb{C}}
\newcommand{\bR}{\mathbb{R}}
\newcommand{\bZ}{\mathbb{Z}}
\newcommand{\bH}{\mathbb{H}}

\newcommand{\fe}{\mathfrak{e}}

\renewcommand{\gg}{\mathfrak{g}}

\newcommand{\gu}{\mathfrak{u}}

\newcommand{\gso}{\mathfrak{so}}
\newcommand{\gsu}{\mathfrak{su}}
\newcommand{\gspin}{\mathfrak{spin}}
\newcommand{\gsp}{\mathfrak{sp}}

\newcommand\Sp[1]{\mathrm{Sp}(#1)}
\newcommand\GL[1]{\mathrm{GL}(#1)}

\newcommand\SO[1]{\mathrm{SO}(#1)}
\newcommand\SU[1]{\mathrm{SU}(#1)}
\newcommand\U[1]{\mathrm{U}(#1)}
\newcommand\Spin[1]{\mathrm{Spin}(#1)}

\newcommand{\cI}{\mathcal{I}}

\def\p{\partial}
\newcommand{\n}{\nabla}

\newcommand\der[1]{\frac{\partial}{\partial #1}}

\newcommand{\ol}{\overline}
\newcommand{\op}{\oplus}

\newcommand{\ra}{\rightarrow}

\newcommand{\ba}{\ol{\a}}
\newcommand{\bb}{\ol{\b}}
\newcommand{\bg}{\ol{\g}}
\newcommand{\bd}{\ol{\d}}
\newcommand{\bh}{\ol{\h}}
\newcommand{\bk}{\ol{\k}}
\newcommand{\br}{\ol{\r}}

\newcommand{\bz}{\ol{z}}

\DeclareMathOperator\Tr{Tr\;}

\DeclareMathOperator\Hol{Hol\,}
\DeclareMathOperator\vol{vol}
\DeclareMathOperator{\Real}{Re}
\DeclareMathOperator{\Imag}{Im}
\DeclareMathOperator{\id}{id}

\makeatletter
\newdimen\rh@wd
\newdimen\rh@hta
\newdimen\rh@htb
\newbox\rh@box
\def\rh@measure#1{\setbox\rh@box=\hbox{$#1$}\rh@wd=\wd\rh@box \rh@hta=\ht\rh@box}

\def\widecheck#1{\rh@measure{#1}%
  \setbox\rh@box=\hbox{$\widehat{\vrule height \rh@hta width\z@ \kern\rh@wd}$}%
  \rh@htb=\ht\rh@box \advance\rh@htb\rh@hta \advance\rh@htb\p@
  \ooalign{$\vrule height \ht\rh@box width\z@ #1$\cr
           \raise\rh@htb\hbox{\scalebox{1}[-1]{\box\rh@box}}\cr}}
\makeatother

\newtheorem{Th}{Theorem}
\newtheorem{Prop}{Proposition}
\newtheorem{Cor}{Corollary}
\newtheorem{Lem}{Lemma}

\newcommand{\bt}{\begin{Th}\ }
\newcommand{\et}{\end{Th}}
\newcommand{\bp}{\begin{Prop}\ }
\newcommand{\ep}{\end{Prop}}
\newcommand{\bc}{\begin{Cor}\ }
\newcommand{\ec}{\end{Cor}}
\newcommand{\bl}{\begin{Lem}\ }
\newcommand{\el}{\end{Lem}}
\newcommand{\ed}{\end{Def}}
\newcommand{\bconj}{\begin{Conj}\ }
\newcommand{\econj}{\end{Conj}}
\newcommand{\pf}{\noindent{\it Proof:\ }}
\newcommand{\qed}{\hfill $\square$}
\newcommand{\half}{\tfrac12}

\newcommand{\be}{\begin{equation}}
\newcommand{\ee}{\end{equation}}
\newcommand\la[1]{\label{#1}}
\newcommand\re[1]{\eqref{#1}}
\newcommand{\arr}{\begin{array}{rlll}}
\newcommand{\ea}{\end{array}}
\newcommand{\bea}{\begin{eqnarray}}
\newcommand{\eea}{\end{eqnarray}}
\newcommand{\bean}{\begin{eqnarray*}}
\newcommand{\eean}{\end{eqnarray*}}

\title{\vskip -3cm
        \begin{flushright}
        \mbox{\normalsize  AEI-2012-207}\\[-12pt]
        \end{flushright}
        \vskip 2cm
       Oxidation of Self-Duality to 12 Dimensions\\ and Beyond
         \vskip 15mm
        \Large   Chandrashekar Devchand \\[8mm]
        \normalsize     Max-Planck-Institut für Gravitationsphysik (Albert-Einstein-Institut)\\
              Am Mühlenberg 1, D-14476 Potsdam, Germany
              }
\date{}
\begin{document}

\maketitle

\begin{abstract}
Using (partial) curvature flows and the transitive action of  subgroups of 
O$(d,\mathbb Z)$ on the indices  $\{1,\dots,d \}$ of the components of the 
Yang-Mills curvature in an orthonormal basis,  we obtain 
a nested system of equations in successively  higher dimensions $d$, each
implying the Yang-Mills equations on $d$-dimensional Riemannian manifolds 
possessing special geometric structures. This `matryoshka' of self-duality equations
contains the familiar self-duality equations  on Riemannian 4-folds as well as 
their generalisations on complex Kähler 3-folds and on  7- and 8-dimensional
manifolds with G$_2$ and Spin(7) holonomy. The matryoshka allows enlargement
(`oxidation') to a remarkable system in 12 dimensions invariant under Sp(3). 
There are hints that the underlying geometry is related to the sextonions,
a six-dimensional algebra between the quaternions and octonions.

\end{abstract}

\newpage
\section{Introduction}

Many interesting examples of special geometric structures on $d$-dimensional Riemannian 
manifolds  $(M,g)$ are provided by certain $G$-invariant covariantly constant (parallel)  
$p$-forms  $\varphi \in \La^pT^*M$, where $G=\Hol$, the restricted holonomy group of $M$.
If  $p < d $,  then  $G$ is clearly a proper subgroup of  SO($d$), since in the 
generic rotationally invariant case, only the volume form is invariant. 

For Riemannian manifolds which are locally neither a product of lower dimensional 
spaces nor a symmetric space, 
Berger's list \cite{B} provides the most interesting examples of restricted holonomy groups. 
These include  $\U{n} \subset \SO{2n}$, which leaves the Kähler two-form $\o$ on a 
$2n$-dimensional Kähler manifold invariant. The $\SU{n}$ Calabi-Yau specialisation has, 
in addition, an invariant complex $n$-form, the holomorphic volume form. 
The group   $\Sp{n} \subset \SO{4n},\, d=4n\,,$ of $n\times n$  matrices with
quaternion elements satisfying $A^\dagger A =1$, 
has three invariant 
Kähler two-forms $\o_\a$,  combinable in a two-form, $\o = \o_1 i + \o_2 j+  \o_3 k$,
taking values in the imaginary quaternions.  These characterise hyper-Kähler geometry. 
The quaternionic Kähler generalisation has $\Hol = \Sp{n}\cdot \Sp{1} \subset \SO{4n}$, 
with the three Kähler forms existing only locally. Globally, they define an invariant parallel
four-form $\sum \o_\a \wedge \o_\a$. The two exceptional $d=7$ and 8 geometries with 
$\Hol = G_2$ and $\Spin{7}$ have, respectively, an invariant three- and four-form. 
In all these cases, the geometric information can equally well be encoded uniformly in
an invariant four-form: the two-forms afford squaring and the three-form in
seven dimensions has a Hodge-dual four-form. 
The Lie group inclusions 
\begin{equation*}
\Sp{n} \subset \SU{2n} \subset \U{2n} \subset   \SO{4n}
\label{}\end{equation*}
imply corresponding inclusions of geometries: hyperkähler manifolds are Calabi-Yau manifolds,
the latter are Kähler, which in turn are orientable. 
The two exceptional cases are also part of  lower dimensional sets of inclusions:
\begin{eqnarray*}
\U{2}  \subset \Sp{2} \subset \SU{4} \subset \Spin{7} \subset   \SO{8}&\\
 \SU{3} \subset G_2 \subset \Spin{7}  \subset  \SO{8}&\hskip -4mm.
\end{eqnarray*}
The respective invariant tensors  can be obtained by successive reductions of the 
$4n$-dimensional volume form. For instance, the Spin(7) invariant four-form
in eight dimensions contracted with an arbitrary vector yields the $G_2$-invariant 
three-form in the orthogonal seven-dimensional space. Similarly, the latter yields
an $\SU{3}$-invariant two-form on projection to the complex three-fold orthogonal 
to an arbitrary vector.

For  Riemannian manifolds $(M,g)$ admitting a $G$-structure, a principle 
subbundle of the frame bundle of $M$, with
structure group $G\subset \GL{d,\bR}$,  the tangent space at every point
admits an isomorphism with $\bR^d$. For every point $p\in M$ there exists
a choice of local coordinates with $p$ as the origin in which the 
Riemannian metric takes  the euclidean form  
$d^2 s = g_{ij} dx^{i}dx^{j} = \sum_i dx^{i}dx^{i}$ 
and the  special geometric structure $\varphi$  in these coordinates
is the constant $G$-invariant form
\begin{equation}
\varphi = \sum_{(i_1,\dots,i_p)\in \cI^+} dx^{i_1,\dots,i_p}\ .
\label{special}\end{equation}
where   $dx^{i_1,\dots,i_p} := dx^{i_1}\wedge \dots\wedge dx^{i_p} $ and
$\cI^+$ is a set of oriented subsets $\{i_1,\dots,i_p\}\,\subset\,
\{1,\dots,d\}$ with $\f_{i_1 \ldots i_p}\,=\,1\,$.
Differential forms  like $\varphi$ have been called {\it special democratic forms} 
\cite{dnw1,dnw2}. They are `special' in the sense that they  have components
$\varphi_{\mu_1\dots\mu_p}$ equal to $+1,-1$ or $0$ in some orthonormal basis,
just like the volume form
$\vol_d =  dx^1\wedge dx^2\wedge\dots\wedge dx^d =: dx^{12\dots d}$
on a Euclidean vector space.  More precisely,  a $p$-form $\f$ is called {\it special} 
if it lies in the $\mathrm{SO}(d,\bR)$-orbit of
 \begin{equation}
  \f\;\;=\;\;\sum_{1\leq \mu_1<\ldots<\mu_p\leq d}
  \f_{\mu_1\ldots\mu_p}\,dx^{\mu_1\ldots\mu_p}
 \la{special1}
 \end{equation}
 with components $\f_{\mu_1\ldots\mu_p}\,\in\,\{-1,0,1\}$.
There are clearly only a finite number of orbits of special $p$-forms 
parametrised by the components $\f_{\mu_1\ldots\mu_p}\,
\in \,\{-1,0,1\}$ under $\mathrm{SO}(d,\bR)$ or $\mathrm{O}(d,\bR)$.
Distinct sets of components may give rise to
special $p$-forms in the same orbit, because the
subgroups $\mathrm{SO}(d,\bZ)\,\subset\,\mathrm{SO}(d,\bR)$ or 
$\mathrm{O}(d,\bZ)\,\subset\,\mathrm{O}(d,\bR)$ 
map the special form $\f$ in equation \re{special}
into a special form parametrised by different components. These groups
are isomorphic to the semidirect product of the permutation group $S_d$
acting naturally on $d{-}1$ or $d$ copies of $\bZ_2$, namely 
$\mathrm{SO}(d,\bZ)\,\cong\,S_d\ltimes\bZ_2^{d-1}$ or
$\mathrm{O}(d,\bZ)\,\cong\,S_d\ltimes\bZ_2^d$. 
Thus, special $p$-forms which appear to be different may nevertheless be in the 
same orbit under $\mathrm{SO}(d,\bR)$ or $\mathrm{O}(d,\bR)$. 
The orbit of a special $p$-form may always be labelled by 
a choice of a representative \re{special}. 

A special $p$-form $\varphi$  is called democratic if its set of
nonzero components $\{\varphi_{i_1\dots i_p}\}$ is symmetric under 
the transitive action of a subgroup of $\mathrm{O}(d,\bZ)$ 
on the indices $\{1,\dots,d\}$. 
The action of an element $(\s,\eta_1,\dots,\eta_d)\in S_d\ltimes\bZ_2^d$,
on the components of $\f$ being given by
\begin{equation}
\f_{i_1\ \dots\ i_p} \mapsto \eta_{i_1}
\dots\eta_{i_p}\,\f_{\s(i_1)\ \dots\ \s(i_p)}\ ,
\label{tsfOd}
\end{equation}
where 
$\eta_i^2=1\,,\,i=1,\dots,d$. 
So for a  democratic form no choice of indices is privileged. 
We refer to \cite{dnw1,dnw2} for further details. It was shown in  \cite{dnw1} that knowledge of the above symmetry groups allows  an enlargement (oxidation) 
of the base space;
the symmetries may be used to remix the sets of indices $\{(i_1\dots i_p)\}$ of the
nonzero components amongst a larger set of indices $\{1,\dots,D\}\,,\, D>d$,
thus defining special democratic $P$-forms in $D$ dimensions from special democratic 
$p$-forms  in  $d$ dimensions for successively higher 
$P \geq p$ and $D \geq d$.  
In this paper, we consider two such oxidation maps\footnote{The notion of using 
the inverse of dimensional reduction as a means of searching for higher dimensional 
parents of lower dimensional theories  goes back to early work on supergravity 
by B. Julia \cite{julia}.}:
\medskip

\noindent
{\bf a) Oxidation through remixing}
\medskip

\noindent
This is a map $\ \varphi\in \Lambda^p\bR^d  \rightarrow  \Lambda^p\bR^D \ni \F\ $
defining a special democratic $p$-form $\F$ 
in  $D> d$ dimensions  in terms of the components of a special  $p$-form $\f$ in 
$d$-dimensions thus:
\begin{equation}
\varphi = \sum_{(i_1,\dots,i_p)\in \cI^+} dx^{i_1,\dots,i_p}  \quad\longmapsto\quad
\F =  \sum_{ \sigma\in H  \subset S_D}  
            \sum_{(i_1,\dots,i_p)\in \cI^+} dx^{\sigma(i_1)\,\dots\, \sigma(i_p)} ,
\la{qb}
\end{equation}
where  $H$ is some subgroup of the symmetric group $S_D$ acting on the $D$ indices.

\bigskip
\noindent
{\bf b) Oxidation through heat flow}
\medskip

\noindent
Alternatively, for $D=d+q$ the nonzero components of a special democratic 
$P=p+q$-form are given by a map 
$\ \varphi\in \Lambda^p\bR^d  \rightarrow  \Lambda^{p+q}\bR^{d+q} \ni \F\ $
defined by
\begin{equation}
\varphi = \sum_{(i_1,\dots,i_p)\in \cI^+} dx^{i_1,\dots,i_p}  \ \longmapsto\ 
\F =  \sum_{ \sigma\in H  \subset S_D}  
            \sum_{(i_1,\dots,i_p)\in \cI^+} 
             dx^{\sigma(i_1)\,\dots\, \sigma(i_p) 
                    \sigma(d+1)\,\dots\, \sigma(D)} .
\la{qc}
\end{equation}

Using these mappings, a nested structure of special forms 
in successively higher dimensions emerges. This is reminiscent of a matryoshka 
({\cyr matr\"eshka}),  a set of nested Russian dolls, traditionally carved in wood, 
where the inner surface of each doll is basically a copy of the outer surface of the 
previous doll; but the outer surface can then vary somewhat, depending on the
geometry of the bulk.

A remarkable nested structure of special democratic forms was displayed in  \cite{dnw1}, 
which included a U(3)-invariant 2-form in six dimensions, a G$_2$-invariant 3-form 
in seven dimensions, and a Spin(7)-invariant 4-form in eight dimensions;
corresponding to the embeddings $\SU{3} \subset G_2 \subset \Spin{7}$ mentioned 
above. It was also shown, that this matryoshka with 3 dolls fits into even larger
dolls and  interesting properties of a  special democratic 6-form in ten dimensions
were presented. 

Motivated by the discussion in \cite{dnw1} of nested special democratic forms, 
we shall presently show that there exists
a corresponding matryoshka of self-duality equations in successively higher dimensions;
each implying the Yang-Mills equations, just as four-dimensional self-duality \cite{cdfn}.
Successive sets of equations are `oxidised' to higher dimensions and `reduced' to lower dimensions 
by enhancing or restricting the permutation symmetries on the sets of indices of
special geometric tensors.   
Remarkably,  the simplest case of the mapping \re{qc}, with $\,q=D-d =1\,$ 
corresponds to equations for (partial) curvature flows for the vector potentials,
hence  `Oxidation through heat flow'.
Solutions of the lower dimensional equations then provide initial values for the 
flow into the extra dimension, the flow to the next doll of the matryoshka.
We shall display oxidations up to $d=16$. The representation theory underlying 
the twelve dimensional system seems to be related to a mathematical curiosity, 
the algebra of the {\it sextonions}  \cite{we,lm}, a six-dimensional algebra between 
quaternions and octonions. This algebra gives rise to a new row in Freudenthal's
magic chart, corresponding to a (non-simple) Lie algebra between 
$\fe_7$ and $\fe_8$, which  has been called $\fe_{7\frac 12}$ 
\cite{lm}. 

\section{Generalised duality for gauge fields in  d $>$ 4}

Generalisations of the four-dimensional self-duality equations to higher dimensions
were introduced some time ago in \cite{cdfn}, where it was shown that restrictions 
of the Yang-Mills curvature two-form $F$ to an eigenspace of a four-form $T$,
implies the Yang-Mills equations. In a standard orthonormal basis of $T^*M$ these
take the form, 
 \begin{equation}
\frac12   g^{km} g^{ln} T_{ijkl} F_{mn} = \lambda F_{ij}, \quad i,j,\dots= 1,\dots,d\ .
\label{tdual}
\end{equation}
Here  $T_{mnpq}$  is a covariantly constant tensor, $g^{pr}$ the inverse metric tensor and
$F=dA+A\wedge A$  is the curvature of a connection $D=d+A$ on a 
Riemannian $d$-fold $(M,g)$ with values in the Lie algebra of a real gauge group 
contained in $\GL{n,\bR}$. 
These  partial-flatness conditions on the curvature  
are first order equations for the vector potentials $A$, so they are more amenable 
to solution than the second order Yang-Mills equations.
Indeed, many special solutions are known (see e.g. \cite{fn,funi, bdn}). 
The  usefulness  of the linear curvature constraints \re{tdual}
follows from the observation \cite{cdfn}:
\bt\la{re_thm} 
For nonzero eigenvalues $\lambda$,  the conditions \re{tdual} imply the 
Yang-Mills equations  $g^{ij} D_i F_{jk}=0$. Thus, potentials $A$ satisfying 
these first order equations automatically  satisfy the Yang-Mills equations.
\et
This result follows in virtue of the Bianchi identities  $\ D_{[i} F_{mn]}\equiv 0\,$. 
In  \cite{cdfn}, constant
four-forms $T$ in flat euclidean spaces were considered, but it is clear that, more
generally  \cite{acd}, it suffices for the consistency condition  
\begin{equation}
 g^{km} g^{ln}(g^{ip}\n_p T_{ijkl}) F_{mn}=0
\la{coclosed}\end{equation}
 to hold, which follows if  
$T$ is co-closed,  $ g^{ip}\n_p T_{ijkl} =0$.  
The latter in turn follows if $T$ is parallel (i.e.\ covariantly constant) with respect to
the Levi-Civita connection  $\n$.  
In dimensions $d> 4$, the four-form $T$ clearly breaks the $d$-dimensional 
rotational invariance
of the Yang-Mills equations. Examples of 4-forms and the corresponding 
partial-flatness conditions \re{tdual} invariant 
under various subgroups   $G \subset \SO{d}$ were studied in \cite{cdfn}
for dimensions $4 < d \le 8$. In particular, interesting examples invariant under
(SU($n$)$\otimes$U($1$))/$\bZ_2$ and  SU($n$), G$_2$ and Spin(7), 
in dimensions $d= 2n, 7, 8$  were constructed.   
The example of $\Sp{n}\otimes\Sp{1}/\bZ_2$ was discussed shortly thereafter 
in \cite{w,cgk}.
The above groups are precisely the holonomy groups of Calabi-Yau, quaternionic Kähler and 
exceptional holonomy manifolds, so remarkably, the generalisations of self-duality for 
most of Berger's special holonomy manifolds \cite{B} were unwittingly constructed before
the subject acquired widespread differential geometric interest (e.g. \cite{uy,cs,n,in,dt,t,acd}).
On all the above manifolds, there exists a $\n$-parallel  four-form,
so the above-mentioned consistency condition on $T$ is satisfied. 

On a $d$-dimensional  Riemannian manifold $(M,g)$  the volume form
$\vol^g$, associated to the metric and orientation, is given in local coordinates
by  $\vol^g = \sqrt{\det(g)} dx^1 \wedge \dots \wedge dx^d$.
The Hodge star operator on the space of $p$-forms  
$* \colon \La^p T^*M \ra \La^{d-p}T^*M$  is defined by  
$\a \wedge * \beta = (\a,\b) \vol^g$,
where $\a,\b \in \La^pT^*M$ and $(\a,\b) $ is the 
natural inner product  induced on $p$-forms by the metric;
 $(\a,\b) = \a_{i_1 \dots i_p}\b_{j_1 \dots j_p} g^{i_1 j_1} \dots g^{i_p j_p} $.
Given the existence of a $G$-invariant four-form $T$ on $M$, the space of two-forms
naturally decomposes into its $T$-eigenspaces,  
$\La^2T^*M = \bigoplus_{\lambda\in \s_T}  V_\l\,$, 
where $\s_T$ is the spectrum of $T$ and the eigenspaces $ V_\l$ are $G$-modules. 
The endomorphism  defined by the four-form $T$ on the space of two-forms is traceless, so   
$\sum_{\lambda\in \s_T}\  d_\l  \lambda =0\ ;\  d_\l := \dim V_\l\,$.

The Yang-Mills equations are the Euler-Lagrange equations for the Yang-Mills functional
defined on the space of vector potentials,
\begin{equation}
S = \| F \|^2   = \langle F , F \rangle   := \int_M  \Tr  (F , F) \vol^g  =  \int_M \Tr  F\wedge *F\ .
\la{ym}\end{equation} 
The topological functional associated with a co-closed four-form $T$ 
yields a Chern-Simons-like form on the boundary of $M$, 
\bea
Q =   \int_M  \Tr  *T \wedge  F\wedge  F 
=  \int_M  \Tr  d \left( *T  \wedge  \left(  A \wedge dA 
                                         + \tfrac23  A\wedge  A\wedge A \right) \right) .
\la{Q}\eea
This functional was used in \cite{bks} to construct examples of higher dimensional
analogues of topological field theories; its relation to the topology of the  bundle 
for some specific examples has been discussed, for instance, in \cite{in,t}.  
The curvature decomposes  into  its $T$-eigenspaces,  
$ F=  \sum_{\lambda\in \s_T}  F_\lambda\, $, 
where the components  $F_\lambda\in V_\l$ satisfy  \re{tdual}
and are mutually orthogonal  with respect to the $L^2$ inner product in \eqref{ym}; i.e.
$ \langle  F_\l , F_\mu  \rangle = 0$ for $\l\neq \mu$. 
Equation \re{tdual} may be expressed in terms of  the projection operator 
$\,P\colon \La^2T^*M \ra \left(V_\l\right)^\perp $ to 
the orthogonal complement of the eigenspace $V_\l$ (see e.g. \cite{dn}),
\begin{equation}
F = F_\l  \quad \Leftrightarrow\quad  P^I_{ij} F^{ij} = 0 \ ,\quad I=1,\dots, \bar d_\l\,,
\la{proj}\end{equation}
where the number of equations, $\bar d_\l\,$, is the codimension of the 
eigenspace $V_\l$ with eigenvalue $\l$.  Here, the projector $P$ is the 
analogue of the 't  Hooft tensor in four dimensions and we lower (raise)
indices  using the (inverse) Riemannian metric. 
The most interesting sets of equations \re{proj}  correspond to the projections 
to the largest eigenspace of $T$; these being the least overdetermined systems. 
We choose the convenient orientation and normalisation of the four-form $T$ 
in which this largest eigenspace has eigenvalue $\l=-1$, corresponding in 
four dimensions to anti-self-duality.

Now, the eigenspace decomposition of $F$, together with eq. \re{tdual}, implies that
\begin{equation*}
*T\wedge F  =  \sum_{\lambda\in \s_T}  *T\wedge F_\lambda 
=  \sum_{\lambda\in \s_T}  \lambda  *F_\lambda\ .
\end{equation*}
So the topological functional in \eqref{Q} has the spectral decomposition
\begin{equation}
Q =  \int_M  \Tr  *T \wedge  F\wedge  F   
=  \sum_{\lambda\in \s_T}  \lambda  *F_\lambda  \wedge F_\lambda  
=   \sum_{\lambda\in \s_T}\  \lambda\   \| F_\lambda \|^2 . 
\end{equation}
Similarly, the Yang-Mills action takes the form
$\ S =  \sum_{\lambda\in \s_T}\   \| F_\lambda \|^2 $. 
Picking out a specific sector of the theory with eigenvalue $\mu$, we may write
\begin{equation}
Q  =  \mu  \| F_\mu  \|^2  
         +  \sum_{\lambda\in \s_T\backslash \{\mu\}}   \lambda\   \| F_\lambda \|^2 .
\end{equation}
It follows that
\begin{equation}
S\ =\   \| F_\mu  \|^2  +  \sum_{\lambda\in \s_T\backslash \{\mu\}}     \| F_\lambda \|^2   
  \ =\  \frac{1}{\mu}  Q  
            +  \sum_{\lambda\in \s_T}  \left( 1- \frac{\lambda}{\mu}\right)  \| F_\lambda \|^2  .
\label{S}\end{equation}
Thus, if $\  ( 1- \frac{\lambda}{\mu})  \ge 0\ $ for all $\lambda\in\s_T$, 
then the action is bounded below:
\begin{equation}
S =  \| F \|^2 \ge   \frac{1}{\mu}  Q \, .
\label{bog}\end{equation}
This $L^2$-bound on the curvature is saturated  precisely when  $F$ is projected 
to the eigenspace with eigenvalue $ \mu$.  Vector potentials satisfying $F(A)=F_\mu(A)$ 
thus minimise the functional $S$, 
with the minimal action taking the value  $\| F_\mu(A) \|^2$. 
The bound \re{bog} shows that  manifolds with a co-closed four-form $T$ differ essentially 
from  spheres. It is known that a connection over  $M= S^d\,,\, d\ge 5$,  with sufficiently
small $L^2$-norm is necessarily flat \cite{na, bls}. This reflects the argument in  \cite{jt}  
that in flat spaces of dimension $d\ge 5$, the only finite action Yang-Mills solutions are 
pure-gauge with $F=0$.  
In the case of more general compact manifolds $M$, curvature 
estimates have been discussed in \cite{p,na2}. Using these results, Tian \cite{t} has 
proven the existence of finite action solutions to \eqref{tdual} on manifolds $M$ with 
a co-closed four-form $T$ satisfying certain conditions, in particular when 
$T$ is a certain calibrating form \cite{hl}. Further, in certain cases, the moduli space 
of equivalence classes of solutions to \eqref{tdual} has also been shown to be 
a compact manifold \cite{t}.

Note that  if $\, a:=  ( 1- \frac{\lambda}{\mu})$ does not have the same sign for 
every $\lambda \neq \mu$, so that the {\em difference} of two sums of squares 
(the terms with $a>0$ and those with $a<0$) gives the action plus a topological term, 
then the equation $F(A)=F_\mu(A)$ determines saddle points of the Yang-Mills functional. 
This yields a theory with no finite lower bound to the action and solutions $A$ 
are necessarily unstable under those infinitesimal deformations $\delta A$ which 
contribute  to curvature components $F_\l$, where  $\l/\mu > 1$. 
For such solutions  the second variation of the action functional
$\delta^2 S / \delta A^2 $ is not non-negative. 
Particular examples of such systems of equations  for saddle points of the Yang-Mills action
have been known for some time \cite{df}.
The quantum field theoretic significance of negative eigenvalues of the second variation
of the action has been discussed for instance in \cite{c,w_morse}.
 
In even dimensions, with  $d=2n$, if the manifold $M$ admits a complex structure $J$,
this provides, at  any point $p$ in $M$,   a linear map  $J_p\colon\ T^*_pM \ra T^*_pM$ 
under which the complexification $T^*_pM \otimes_\bR \bC$ 
splits into the eigenspaces $T_p^{(1,0)}M$  and  $T_p^{(0,1)}M$, both of which are 
isomorphic to $\bC^n$. This allows the choice of complex coordinates $(z^1,\dots,z^n)$ 
and $(z^{\bar 1},\dots,z^{\bar n})$. 
The complex (1,0)- and (0,1)-forms 
$\{  dz^\a\}$ and  $\{dz^{\ba}\}$, for $\a,\ba=1,\dots,n$, 
then provide bases for $T_p^{(1,0)}M$  and  $T_p^{(0,1)}M$ respectively.
Imposing the reality conditions $dz^{\ba} = \ol {dz^\a}$, we may 
recover  $ \bR^{2n} \simeq \bC^n $. 
The  curvature two-form in this basis has components 
$F_{\a\b}, F_{\a\bb}, F_{\ba\bb} = \ol F_{\a\b} $ and
the Riemannian metric locally takes the hermitian form  
$d^2 s = g_{\a\bb} dz^{\a}dz^{\bb} = dz^{\a}dz_{\a} =\sum_\a dz^{\a}dz^{\ba}$ 
and the  complex $(n,0)$ volume form is given by  
$\O =  dz^{\a_1\dots\a_n}$.
In the complex setting, equation \re{tdual} is a $G$-invariant
equation, where the structure group $G$ is a subgroup of 
$\GL{n,\bC} \subset \GL{2n,\bR}$.   
For the particularly important $\l =-1$ case, we shall use the following
complex variant of Theorem \ref{re_thm}.

\bt \la{cx_thm}
On a Riemannian complex $n$-fold $(M^{2n},g)$,  with hermitian metric
$g= g_{\a\bb} dz^\a dz^{\bb} $ and (4,0)-form $\F$,  the linear curvature constraints,
\begin{eqnarray}
 F_{\a\b}  + \half  g^{\g\bh}  g^{\d\bk}\  \F_{\a\b\g\d}\  F_{\bh\bk}  &=& 0\,,    \la{hol}\\   
g^{\a\bb} F_{\a\bb}  &=& 0\,,     \la{re}   \\
 g^{\g\bh}  g^{\d\bk}\  (g^{\a\br} \n_{\br} \F_{\a\b\g\d})\  F_{\bh\bk}  &=& 0\,, \la{torfree}
\end{eqnarray}
imply the Yang-Mills equations  
$ g^{\a\br} D_{\br} F_{\a\b} =  g^{\a\br} D_{\br} F_{\a\bb} =0$.
\et
\pf Using \re{hol} we have
\begin{equation}
 g^{\a\br} D_{\br} F_{\a\b} = D^{\a} F_{\a\b}=  - \half  (\n^{\a} \F_{\a\b\g\d})\  F^{\g\d} 
                               - \half   \F_{\a\b\g\d}\ D^{\a} F^{\g\d} = 0,
\label{pf1}\end{equation}
the first term being the left side of \re{torfree} and the second vanishes in virtue
of the Bianchi identity $\ D_{\ba} F_{\bg\bd} +\ {\rm cyclic\ permutations}\ = 0$.
Similarly, using the  Bianchi identity between  $D_{\br},D_{\bb}$ and $D_{a}$ we have,
$
D_{\br} F_{\a\bb} =   D_{\bb} F_{\a\br} +  D_{\a} F_{\ba\br}\ .
$
On contracting with $g^{\a\br}$, the second term on the right yields the complex conjugate 
of the left side of \re{pf1} and the
first term contains the trace of the (1,1)-part of the curvature, which vanishes by 
equation \re{re}. 
\qed

Already in \cite{cdfn}, it was noticed that the  lower dimensional cases, including
four-dimensional self-duality, the six-dimensional 
SU($3$)$\otimes$U($1$))/$\bZ_2$-invariant equations and the seven-dimensional  
G$_2$-invariant equations,  were reductions of the eight-dimensional 
Spin(7)-invariant set of equations. In the present paper, we show that 
using the results of \cite{dnw1} these equations also admit a systematic  
`oxidation' to higher dimensions starting from the lower dimensional ones. 

We consider two types of oxidation. 
The first is based on the map \re{qb} and uses cyclic permutations to remix the index sets appearing in the lower dimensional equations  amongst a larger set of indices.  
The second oxidation method is based on the heat flow for some appropriate partial
curvature. This is  related to the $\,D-d=1\,$ case of \re{qc}.  More specifically,
if in $(d{-}1)$-dimensions,  there exist a special set of $d{-}1$ curvature constraints
$f_{ijk} F^{jk} =0\ ,\  i = 1,\dots,d{-}1 $, where $f$ is some appropriate tensor,
 then we can consider the  corresponding 
partial curvature flow 
\begin{equation}
\dot A_i = f_{ijk} F^{jk} \ ,\  i = 1,\dots,d \,.
\la{hfd} 
\end{equation} 
Identifying the parameter of the variation or `time' of the flow with a $d$-th independent variable $x^{d}$,
the left hand side is the  $A_{d}=0$ `temporal' gauge form of the curvature components
$F_{di}$, so that the flow equations \re{hfd} are in fact  linear curvature constraints 
of the form 
\begin{equation}
F_{di}  = f_{ijk} F^{jk}\,.
\label{hfd_cov}\end{equation}
Remarkably, in many interesting cases, these constraints
may be reformulated in the  form \re{tdual}, thus implying the Yang-Mills equations.
The idea of choosing such a temporal gauge to obtain a flow equation is not new.
For instance, both Nahm's equations for magnetic monopoles \cite{cg} and the 
generalisations to higher dimensions of Euler's equations for a spinning top \cite{fu}, 
arise from the  imposition of precisely such a gauge choice on equations 
of the form \re{tdual}.  Conversely, the flow equations can be written 
as gauge covariant equations in one dimension higher by gauge un-fixing the 
component of the gauge potential in the direction of the flow.
This idea has also been used by Tao \cite{tao} in the context of the 
full second order Yang-Mills  gradient flow  $\dot A_k = g^{ij}D_iF_{jk}$. 

As we shall see, the juxtaposition of the  two oxidation methods above yields 
the advertised matryoshka of self-duality equations, starting from zero curvature
in $d=2$ and including the familiar 4-dimensional self-duality, as well as its
generalisations to 6,7 and 8 dimensions mentioned above. Remarkably,
the matryoshka affords enlargement to even higher dimensions. 
We discuss an interesting 12 dimensional extension and display its oxidation 
to 14 and 16 dimensions.

\section{The matryoshka of self-duality equations}

Let us begin in  two dimensions 
with the  flatness condition $F_{12}  = 0$ for the sole component of the 
curvature two-form.  In  the complex setting, the curvature only has a (1,1)-part,
$F_{z\bar z}$, where we  use complex coordinates
 $z=x^1 + i x^2\ ,\  \bar z = x^1 - i x^2\ $.
The flatness condition means that the curvature is in the kernel 
of the volume form. We therefore have, 
\begin{equation}
 \e_{ij} F^{ij} = 0   \quad\Leftrightarrow \quad  
F_{12}  = 0 
   \quad\Leftrightarrow \quad  F_{z\bar z} =0\ ,
\label{d2}\end{equation}
Both real and complex forms of the equations  are locally rotationally invariant, 
since their respective invariance algebras $\gso(2)$ and $\gu(1)$ are isomorphic.
The rich properties of the solutions of these equations on Riemann surfaces have
been investigated by Atiyah and Bott \cite{ab}.

We oxidise the equation  $F_{12}  = 0$ to a  system in three dimensions  by  
acting on the indices by all permutations  generated by the  
cycle $\s=(1\ 2\ 3) \in S_3\,$, so as to obtain a system of equations invariant
under these permutations:
\begin{equation}
\{  F_{12}  = 0 \}   \longrightarrow \{  F_{\s^p(1)\s^p(2)}   = 0\ ;\  \s=(1\ 2\ 3) \ ,\ p=1,2 \} 
\label{d2to3}\end{equation}
This of course yields flatness  in 3 dimensions; the curvature lies in the kernel
of the three-dimensional volume form,
\begin{equation}
  \e_{ijk} F^{jk} = 0   \quad\Leftrightarrow \quad  
F_{12} =  F_{23} =  F_{31} = 0\ .
\label{d3}\end{equation} 
Since this is a set of 3 equations for the three vector potentials $A_i\,,\, i=1,2,3$,
it allows us to write the Yang-Mills curvature flow 
\begin{equation}
\der{x^4} A_i (x^i, x^4)  =  \half \e_{ijk} F^{jk}\,,\quad i=1,2,3 \ ,
\la{hf3}\end{equation} 
with initial  (at $x^4 =0$)  flat connection $A_i (x^i, 0)$ satisfying \re{d3}. 
This evolution equation is the gradient flow of the Chern-Simons functional \cite{witten_cs} 
on the space of connections
\begin{equation}
S_{CS} =  \int_{M^3} \Tr (\half AdA +  \tfrac13 A^3) 
=  \int_{M^3} \Tr (\half A_i \p_j A_k + \tfrac13 A_i A_j A_k) dx^{ijk} ,
\label{cs}\end{equation}
where $dx^{ijk} = dx^i\wedge dx^j \wedge dx^k$, the volume form.
In his canonical quantisation of this theory, Witten \cite{witten_cs} considered
the 3-fold to be of the form $M^3= \Sigma \times \bR^1$, where the data on the
2-dimensional boundary $\Sigma$, a Riemann surface,  satisfied the equations
\re{d2}.  

Now applying an $x^4$-dependent gauge transformation to the vector potentials
\begin{equation}
A_a(x^i, 0)\mapsto
 g^{-1}(x^i, x^4) A_a(x^i, 0)  g(x^i, x^4) + g^{-1}(x^i, x^4) \p_a  g(x^i, x^4)\ ,\ a= 1,\dots, 4\ ,
\label{gt}\end{equation}
where $A_4(x^i, 0) =0$,  yields a pure-gauge form for the fourth vector potential, 
$A_4= g^{-1} \p_4 g\,.$
 The non-gauge covariant equation \re{hf3} now takes the  gauge covariant form of the 
four dimensional SO(4)-invariant anti-self-duality equations
\begin{equation}
 F_{ab} + \half \e_{abcd} F^{cd} = 0\,,\quad   a,b,c,d=1,\dots 4\ ,
\la{d4}\end{equation} 
a set of 3 equations for the 4 vector potentials. (The self-duality equations
emerge on reversing the $x^4$-direction of the flow.) 

Using a manifestly $\gu(2)$-covariant notation for Yang's complex coordinates 
$(z^{\a}\ ,\ z^{\ba} := \ol{z^{\a}} \,,\, \a,\ba = 1,2)$, 
these equations take the form (c.f. \re{d2}) \cite{yang}, 
\begin{eqnarray}
 \O_{\a\b} F^{\a\b} &=& 0 \quad\Leftrightarrow \quad  
 F_{\bar 1\bar2}= 0 
  \la{hol4}\\   
g^{\a\bb} F_{\a\bb} &=& 0\quad\Leftrightarrow \quad  
 F_{1\bar1} + F_{2\bar2}= 0 
\ .\la{re4}   
\end{eqnarray}
This is a system consisting of one complex and one real equation, leaving 
as the sole non-zero part, the trace-free part of the (1,1)-curvature.
The U(2)-invariant metric on $\bC^2\simeq \bR^4$ is 
given by $ g_{\a\bb} dz^{\a} dz^{\bb} = dz^{1} dz^{\bar1} + dz^{2} dz^{\bar2} 
=:  dz^{\a} dz^{\ba}\,$ and the symplectic (2,0)  volume form, invariant under SU(2), 
by  $\O= \O_{\a\b} dz^{\a}\wedge dz^{\b} = dz^{1}\wedge dz^{2} =: dz^{12}$.

Now, complexifying all the data by dropping all reality conditions 
(see for instance the discussion in \cite{acd}), we obtain 
the additional equation $F_{\a\b}=0$, which allows us to choose  
the holomorphic gauge $A_{\a}  =0$. The equation \re{re4} then 
takes the form of a conservation law \cite{bfny}
\begin{equation}
  g^{\a\bb} \p_{\a} A_{\bb} =  \p^{\bb} A_{\bb} = 0\ ,\ \  \a,\bb=1,2\,,\la{curr4}   
\end{equation}
which has local solution 
$\ A_{\bb} =   \O_{\bb\bg}  \p^{\bg} f \ $,
where  $\ol\O= \O_{\ba\bb} dz^{\ba}\wedge dz^{\bb} 
= dz^{\bar1 \bar2} $  is the symplectic 
(0,2)-form. 
The remaining equation in \re{hol4} then takes
the form of Leznov's  wave equation  \cite{l}
\begin{equation}
 \square f  + \half   \O^{\a\b}\, [\p_{\a} f\ ,\  \p_{\b} f ] =0\ ,
\label{l}\end{equation}
with Laplacian $\square =   g^{\a \bb}  \p_{\a}\p_{\bb} =  \p^{\ba}\p_{\ba} $.
Solutions provide stationary points of the Leznov functional
\begin{equation}
 S_L = \int_{M^\bC} \Tr \left( \frac12 f \square f
             + \frac13   
             \O^{\a\b} f \p_{\a} f \p_{\b} f \right)  \ .
\label{leznov}\end{equation} 
whose gradient flow has the standard heat equation form, 
\begin{equation}
 \der{t} f  = \square f + \half   \O^{\a\b}\,  [\p_{\a} f\ ,\  \p_{\b} f]\ .
\label{Lflow}\end{equation} 
Here $t$ is the parameter of the infinitesimal variation, so stationary points
correspond to functions $f(t,z,\bz)$ independent of $t$, i.e.\ solutions of eq.\,\re{l}.
In the case of eq.\,\re{Lflow}, the left-hand-side  does not allow interpretation
as a (gauge-fixed) component of the curvature.

In all the above cases, in dimensions $d=1,\dots,4$,  the equations are fully 
SO($d$)-invariant. The special geometric structures characterising these equations
are  thus precisely the volume forms, which are trivially special democratic forms. 
The oxidised volume form in $d$-dimensions $\vol_d= dx^{1\dots d}$ may be obtained from lower
dimensional volume forms by taking successive wedge products with the additional
basis one-forms,   $\vol_d = \vol_{d-1}\wedge dx^d $.

\section{From four to eight dimensions}

\subsection{Permutation to d=6}
\label{sect_d6}
To proceed to higher dimensions, we now consider the complex version 
\re{hol4},\re{re4}
of the four-dimensional equations. 
Following the previous  mapping from two to three dimensions 
\re{d2to3}, we now oxidise these equations  from $\bC^2$ to  $\bC^3$ by 
requiring invariance under the cyclic permutations generated by  
$\s=(1\ 2\ 3) \in S_3\,$, where  the indices are now complex;
\begin{equation}
\{  F_{\bar 1\bar2} = 0 \}   \longrightarrow \{  F_{\s^p(\bar 1)\s^p(\bar 2)}   = 0\ ;\ 
 \s=(1\ 2\ 3) \ ,\ p=1,2 \}.
\label{c2toc3}\end{equation}
This yields the system  (c.f. \re{d3})
\begin{eqnarray}
\O_{\a\b\g} F^{\b\g} = 0   &\Leftrightarrow&
\{ F_{\bar 1\bar2}=F_{\bar 2\bar3}=F_{\bar 3\bar1}=  0 \}  
 \la{hol6}\\[4pt]   
g^{\a\bb} F_{\a\bb} = 0
&\Leftrightarrow&
 F_{1\bar1} + F_{2\bar2}  + F_{3\bar3}= 0 \  ,\ \  \a,\ba=1,2,3\ ,\la{re6}   
\end{eqnarray}
a set of three complex and one real equation.
Here $ g_{\a\bb} dz^{\a} dz^{\bb}$ is the U(3)-invariant  hermitian metric and 
$\O= dz^1 \wedge  dz^2 \wedge  dz^3 = dz^{123} $, the complex (3,0) volume form. 
These equations were  obtained in \cite{cdfn} as  
SU(3)$\otimes$U(1))/$\bZ_2$-invariant
curvature constraints which imply the second order
Yang-Mills equations.  They later made an appearance in work by Donaldson \cite{donaldson},
Uhlenbeck and Yau \cite{uy} as the equations for holomorphic connections on 
three (complex) dimensional Kähler manifolds, $g$ being the Kähler metric.

In the six real coordinates,  $x^\a := \Real z^\a\,,\, x^{\a+3}= \Imag z^\a\,,\, \a=1,2,3$,
the equations take the form \re{tdual}, with the special
democratic four-form  (see \cite{cdfn})
\begin{equation}
T_{(6)}= dx^{1425} +dx^{1436} +dx^{2536}.
\label{t6}\end{equation}
This is invariant under the group $S_3$ of permutations of the 3 ordered pairs  
$(\{1,4\},\{2,3\},\{4,5\})$, or, equivalently, the symmetries generated by the 
permutation $\s = (123)(456) \in S_6$. The stabiliser of $T_{(6)}$ in SO(6) is the group 
$\SU{3}\times\U{1}/\bZ_2$ and under this, the space of 2-forms has the following 
decomposition into 
eigenspaces of $T_{(6)}$   \cite{cdfn}:
\begin{equation}
\La^2{\bR^6} =  \big(\gsu(3)_0\,,\,\l=-1\big) 
\op  \big(V_{2}^3 \op \ol{V}_{-2}^3\,,\,\l=1\big)  
\op \big( \bR\o_0 \,,\,\l=2\big)\,,
\label{La2u3}\end{equation}
where $(V_q^n\,,\,\l) $ is the $n$-dimensional irreducible representation of SU(3),
the index $q$ denotes the U(1) charge, $\l$ the eigenvalue of $T_{(6)}$
and $\o_0 = g_{\a\bb} dz^\a \wedge dz^{\bb}$ is the invariant metric form 
associated with $g$. Two-forms parallel to $\o_0$ are contained in 
the $\l= 2$ eigenspace. 
Under the action of  $T_{(6)}$  the curvature tensor therefore decomposes 
into $T_{(6)}$-eigenspaces according to
\begin{equation}
F =  \big(  F_{\g\bd} - \tfrac13 g_{\g\bd} F_0  \,,\,\l{=}-1\big) 
\op  \big( F_{\a\b} \op F_{\ba\bb}  \,,\,\l{=}1\big)  
\op \big(  F_0 \,,\,\l{=}2\big),
\label{d6F}\end{equation}
where $F_0$ denotes the trace $g^{\a\bb} F_{\a\bb}$.
The set of seven equations \re{hol6}, \re{re6} thus projects the curvature to the 
8-dimensional $\gsu(3)$ part, the $\l=-1$  eigenspace. 
We note that solutions of the set of nine equations projecting to the six-dimensional 
$\l=1$  eigenspace are saddle points of the Yang-Mills action in accordance with
eq.\,\eqref{S}.

Analogously to \re{re4}, complexifying the Yang-Mills fields, 
equation  \re{re6}, in the holomorphic gauge 
$A_{\a} =0\,,\,\a=1,2,3\,,$   can be locally solved in
terms of three prepotentials taking values in the complexification of the gauge group: 
\begin{equation}
A_{\ba} =  \O_{\ba\bb\bg}   \p^{\bg} f^{\bb}\ .
\label{p6}\end{equation}
The remaining conditions \re{hol6} provide extrema of the Chern-Simons action 
\begin{eqnarray}
S &=&  \int_{M_\bC} \Tr (\bar A\bar \p \bar A + \bar A^3) \wedge \ast \ol\O 
\nonumber \\
&=&  \int_{M_\bC}  \Tr  (\half A_{\ba} \p_{\bb} A_{\bg} 
                          + \tfrac13 A_{\ba} A_{\bb} A_{\bg})\ dz^{\ba\bb\bg}\,. 
\label{cs6}\end{eqnarray}
Inserting \re{p6} in \re{hol6} yields a wave equation analogous to \re{l}
for the triplet of complex prepotentials $f_\b$,
\begin{equation}
\p^{\b}  \p_{[\a} f_{\b]} 
+ \half  \O^{\b\d\h} [\p_{\d} f_{\h}\ ,\  \p_{[\a} f_{\b]}] =0\ .
\end{equation}
The associated gradient flow for the functional \eqref{cs6} takes the heat 
equation form,
\begin{equation}
\der{t} f_\a  =   \p^{\b}  \p_{[\a} f_{\b]} 
+ \half \O^{\b\d\h}\ [\p_{\d} f_{\h}\ ,\  \p_{[\a} f_{\b]}]\ .
\label{l6}\end{equation}

The reduction of \re{hol6},\re{re6} to the missing $d=5$ case involves choosing a constant unit vector 
in $\bR^6$ and projecting to the five-dimensional space orthogonal to it. Without loss 
of generality, we may  simply choose one of the basis vectors, say $e_6$,
effectively deleting  the variables $x^6$ and  yielding 
an SO(4)-invariant 4-form $T= dx^{1245}$. The corresponding equations (see \cite{cdfn}) 
are an embedding of four-dimensional self-duality  \re{d4} in five dimensional space.
A five dimensional reduction of the Chern-Simons action \re{cs6} and  corresponding 
flow equations were discussed some time ago by Nair and Schiff \cite{ns}.

\subsection{Flow to d=7 and d=8}

Since the three complex equations \re{hol6} have an action \re{cs6}, we may write down
the partial curvature flow, for the three complex potentials $A_{\a}$, now depending on seven variables 
$(z^{\a}, z^{\ba}, x^7)$:
\begin{equation}
\der{x^7} A_{\a}   =   \O_{\a\b\g}  F^{\b\g} \  ,\ \  \a=1,2,3. \la{hf7}
\end{equation} 
This being the gradient flow for the functional \re{cs6}.
Now, analogously to the four-dimensional case \re{gt}, an $x^7$-dependent gauge transformation 
yields the fully gauge covariant form of this partial curvature flow
\begin{equation}
F_{7\a}   =  \O_{\a\b\g}  F^{\b\g}
\quad\Leftrightarrow\quad
 \{F_{71}   =   F_{\bar 2\bar3}\ ,\  F_{72}   =  F_{\bar 3\bar1}\ ,\   
F_{73}   =  F_{\bar 1\bar2} \}  \ .
\la{hol7}\end{equation} 
Here  $\p/\p x^7 $ denotes the real vector field (the `time' of the flow) and
$ \a,\b,\g= 1,2,3$ are complex indices.  
The three complex equations \re{hol7} together with the real equation,
\begin{equation}
g^{\a\bb} F_{\a\bb} =   F_{1\bar1} + F_{2\bar2}  + F_{3\bar3}= 0 \ , 
\label{re7}\end{equation}
imply the Yang-Mills equations in seven dimensions.
Choosing real coordinates $(x^1,\dots,x^7)$, these equations  they take the manifestly 
G$_2$-invariant  form \cite{cdfn}
\begin{equation}
\psi_{ijk} F^{jk} = 0  \ ,\quad i,j,k=1,\dots,7\,.
\label{d7}\end{equation}
Here $\psi$ is the  G$_2$-invariant Cayley three form whose components $\psi_{ijk}$ provide 
structure constants of the algebra of imaginary octonions. 
Choosing the first six real coordinates as the real and imaginary parts
of the complex coordinates as follows, $\ z^\a = x^\a +i x^{\a +3}\,,\ \a=1,2,3$,
we obtain,
 \begin{equation}
\psi = dx^{367} +dx^{257} +dx^{147} 
 + dx^{465} + dx^{243} + dx^{135} +dx^{162}.
\label{psi}\end{equation}
Its four-form dual is given by
\begin{equation}
\varphi := \ast \psi = dx^{1245} +dx^{1346} +dx^{2356} 
 + dx^{7123} + dx^{1567} + dx^{7246} +dx^{3457},
\label{Tg2}\end{equation}
in terms of which the equations \re{d7} take the form, 
\begin{equation}
F_{ij} + \tfrac12 \f_{ijkl}  F^{kl} = 0  \ ,\quad i,j,k=1,\dots,7\,,
\label{d7b}\end{equation}
which projects the curvature to the  $\l=-1$ eigenspace of  $\varphi$;
the eigenspace decomposition of the space of 2-forms being  \cite{cdfn}
\begin{equation}
\La^2 \bR^7 = (\gg_2 \,,\, \l=-1) \oplus (\bR^7 \,,\, \l = 3).
\label{La2g2}\end{equation}

Since the system \re{d7} consists of 7 equations for 7 potentials
and has the Chern-Simons type action
\begin{equation}
S_{CS} =  \int_{M^7}  \Tr (AdA + A^3) \wedge \ast \psi
=  \int_{M^7} \Tr  (\half A_i \p_j A_k + \tfrac13 A_i A_j A_k)\ \psi^{ijk}\,, 
\label{cs7}\end{equation}
we can immediately write  down the
corresponding partial curvature flow in eight dimensions analogous to  \re{hf3}:
\begin{equation}
\der{x^8} A_i  = \half \psi_{ijk} F^{jk}\,,\quad i=1,\dots,7\, .
\la{hf8}\end{equation} 
This is the temporal gauge ($A_8=0$) form of the Spin(7)-invariant equations in eight dimensions,
which were discovered in  \cite{cdfn} and shown there to arise as the projection of the
curvature form to the $\l=-1$ eigenspace of the Spin(7)-invariant 4-form $\phi$, 
\begin{equation}
 F_{ab} + \half \phi_{abcd} F^{cd} = 0\,,\quad   a,b,c,d=1,\dots 8\ ,,
\la{d8}\end{equation} 
where in terms of the seven dimensional  forms $\psi \,,\, \f$ in \re{d7} and  \re{d7b}
the four-form $\phi$ in eight dimensions is given by
$\phi = dx^8 \wedge \psi + \varphi $. 
The decomposition $\La^2 \bR^8$  into eigenspaces of this 4-form is given by  
\begin{equation}
\La^2 \bR^8 = (\gspin_7 \,,\, \l=-1) \oplus (\bR^7 \,,\, \l = 3).
\label{La2spin7}\end{equation}

In complex coordinates, $z^\a = x^\a +i x^{\a +4}\,,\ \a=1,2,3,4$,   the equations
\re{d8} take the form 
(see  \cite{cdfn}) incorporating \re{hol7},
\begin{eqnarray}
F_{\a\b}   +  \half \O_{\a\b\g\d}\ F^{\g\d} =0
&\Leftrightarrow&
\{F_{41}   =   F_{\bar 2\bar3}\ ,\  F_{42}   =  F_{\bar 3\bar1}\ ,\   
F_{43}   =  F_{\bar 1\bar2} \}  
\nonumber\\[4pt]   
g^{\a\bb} F_{\a\bb}  = 0
&\Leftrightarrow&
 F_{1\bar1} + F_{2\bar2}  + F_{3\bar3} + F_{4\bar4} = 0\ ,\la{d8c}   
\end{eqnarray}
where $g$ is the U(4)-invariant hermitian metric on $\bC^4 \simeq \bR^8$ 
and $\O=dz^{1234}$ is the SU(4)-invariant volume form in $\bC^4$. 
In the complex `temporal' gauge, $A_{4}=0$, the three complex 
equations in \re{d8c} therefore take the form of
a partial curvature flow with complex flow parameter $z^{4}$,
\begin{eqnarray}
  \der{z^{4}} A_{\a}  &=&   \half   \O_{\a\b\g}\  F^{\b\g} \la{hf8cx}  \\[4pt]
  \der{z^{4}} A_{\bar4}  &=&  g_3^{\a\bb} F_{\a\bb} 
\la{hf8re} \end{eqnarray}
where $ \O_{\a\b\g}, g_3^{\a\bb}$ are the volume form and inverse metric
of the complex 3-space  
orthogonal to the complex vector field $\p/\p z^4$.  The equation \re{hf8cx} 
thus gives the complex variation of the Chern-Simons action \re{cs6}.

All the above duality equations in dimensions up to eight are more or less
well-known \cite{cdfn}. Our main result is that the pattern of successive
dimensional oxidation actually continues to higher dimensions. Proceeding
further, we see that a particularly interesting 12-dimensional system results.

\section{Self-duality in 12 dimensions}
Following the method of oxidising the duality equations from $\bR^4$ to  
$\bR^6$, we now extend the 
system \re{d8c} in $\bC^4$  to  $\bC^6$ by juxtaposing two additional complex variables 
$z^5 , z^6$ and then remixing the six complex indices by requiring symmetry under 
permutations generated by  $\s = (135)(246) \in S_6$.   We  thus obtain 
the equations,
\begin{equation}
g^{\a\bb} F_{\a\bb}  =
F_{1\bar1} + F_{2\bar2} + F_{3\bar3} + F_{4\bar4} + F_{5\bar5} + F_{6\bar6} = 0 \ ,
\la{re12}  
\end{equation}
together with
\begin{eqnarray}
F_{12} + F_{\bar3 \bar4} + F_{\bar5 \bar6} = 0\,,& 
F_{34} + F_{\bar5 \bar6} + F_{\bar1 \bar2} = 0 \,,\quad  
F_{56}\ +\!\!  &F_{\bar1 \bar2} + F_{\bar3 \bar4} = 0 
\label{hol12a}\\[5pt]
 F_{13} + F_{\bar4 \bar2} = 0 \,,&
\hskip1mm F_{14} + F_{\bar2 \bar3} = 0 \,,\quad  
 &F_{15} + F_{\bar6 \bar2} = 0  \nonumber\\ 
 F_{16} + F_{\bar2 \bar5} = 0 \ ,&  
\hskip1mm  F_{35} + F_{\bar6 \bar4} = 0 \,,\quad   
 &F_{36} + F_{\bar4 \bar5} = 0 \, .
\label{hol12}
\end{eqnarray}
These equations  imply the 12-dimensional Yang-Mills equations!
The proof follows from Theorem \ref{cx_thm} and the observation that
these equations allow expression in the form  \re{hol}, \re{re}, with 
the (4,0)-form $\F$ taking the form
\begin{equation}
\F = dz^{1234} + dz^{1256} + dz^{3456}\,.
\label{omega12c}\end{equation}
This four-form is thus given by  $\F = \o^2$, where $\o$ is the symplectic (2,0)-form
\begin{equation}
 \o = dz^{12} + dz^{34} + dz^{56} \in \La^2 \bC^6\,.
\label{sympl}\end{equation} 
This is analogous to the $\bR^6$ case, 
except that now everything is complex.  The (4,0)-form $\F$ is manifestly
invariant under the action of $\Sp{3}\subset \SU{6} \subset \Spin{12}$.

The three conditions in \re{hol12a} are equivalent to the four real equations,
\begin{eqnarray}
 \Imag(F_{\bar1\bar2}) =  \Imag(F_{\bar3\bar4}) =   \Imag(F_{\bar5\bar6}) &=&  0    \ ,\nonumber\\ 
   \Real (\o^{\ba\bb}F_{\ba\bb}) = \Real(F_{\bar1\bar2} +   F_{\bar3\bar4} +  F_{\bar5\bar6})&=& 0\,,
\label{real4}\end{eqnarray}
where the symplectic (0,2)-form
$\ol\o = \o_{\ba\bb} dz^{\ba}\wedge dz^{\bb}  
=  dz^{\bar1\bar2} + dz^{\bar3\bar4} + dz^{\bar5\bar6}$ and 
$ \o^{\ba\bb} \o_{\bb\bg}  = \d^{\ba}_{\bg}\,$.
The system of equations thus consists of 5 real equations, \re{re12} and \re{real4},
together with 6 complex equations \re{hol12}, a total of 17 real equations.

The entire system \re{re12},\re{hol12a},\re{hol12} in real coordinates for 
12-dimensional euclidean space given by
$  x^i = \Real z^i \,,\   x^{i+6} = \Imag z^i \,,   i = 1,\dots,6\,,$  takes the following form.
Here  we denote the indices 10,11,12 by  $0,a,b$ respectively.
\begin{eqnarray}
&&F_{12} + F_{34} + F_{56} + F_{87} + F_{09} + F_{ba}  = 0    \label{re12a}\\[5pt]
&&F_{17} + F_{28} + F_{39} + F_{40} + F_{5a} + F_{6b}  = 0    \label{re12b}\\[5pt]
&&F_{13} + F_{42} + F_{97} + F_{80}  = 0  \nonumber \\
&&F_{14} + F_{23} + F_{07} + F_{98}  = 0  \nonumber \\
&&F_{15} + F_{62} + F_{a7} + F_{8b}  = 0  \nonumber \\
&&F_{16} + F_{25} + F_{b7} + F_{a8}  = 0  \nonumber \\
&&F_{35} + F_{64} + F_{a9} + F_{0b}  = 0  \nonumber \\
&&F_{36} + F_{45} + F_{b9} + F_{a0}  = 0  \nonumber \\
&&F_{19} + F_{73} + F_{84} + F_{20}  = 0  \nonumber \\
&&F_{10} + F_{74} + F_{92} + F_{38}  = 0  \nonumber \\
&&F_{1a} + F_{75} + F_{86} + F_{2b}  = 0  \nonumber \\
&&F_{1b} + F_{76} + F_{a2} + F_{58}  = 0  \nonumber \\
&&F_{3a} + F_{95} + F_{06} + F_{4b}  = 0  \nonumber \\
&&F_{3b} + F_{96} + F_{a4} + F_{50}  = 0  \label{re12d} \\[5pt]
&&F_{18} + F_{72}  = 0\nonumber \\ 
&&F_{30} + F_{94}  = 0\nonumber \\ 
&&F_{5b} + F_{a6}  = 0   \label{re12c}
\end{eqnarray}
These equations  have the familiar form \re{tdual}, with the 4-form 
$T_{(12)}\in \La^4\bR^{12}$ 
being given by the special democratic form
\begin{eqnarray}
T_{(12)} &=&\phantom{+}  dx^{1234} + dx^{1256} + dx^{1287} + dx^{1209} + dx^{12ba} + dx^{1397} + dx^{1380} \nonumber\\ &&
+ dx^{1407} + dx^{1498} + dx^{15a7} + dx^{ 158b} + dx^{16b7} + dx^{16a8} + dx^{2307} \nonumber\\ &&
+ dx^{2398} + dx^{ 2479} + dx^{2408} + dx^{25b7} + dx^{25a8} + dx^{267a} + dx^{26b8} \nonumber\\ &&
+ dx^{3456} + dx^{3487} + dx^{3409} + dx^{34ba} + dx^{35a9} + dx^{35b0} + dx^{36b9} \nonumber\\ &&
+ dx^{36a0} + dx^{45b9} + dx^{45a0}  + dx^{469a} + dx^{46b0} + dx^{5687} + dx^{5609}\nonumber\\  &&
+ dx^{56ba} + dx^{7890} + dx^{78ab}  + dx^{90ab},
\label{tt12}\end{eqnarray}
which has  a set of 39 non-zero components.
The characteristic polynomial of this Sp(3)-invariant four-form, acting on the 
space of two-forms has been calculated using Maple. 
The eigenspace decomposition of the space of 2-forms in terms of Sp(3)
representations  (see e.g. \cite{MP,ov}) is given by
\begin{eqnarray}
\Lambda^2\bR^{12} 
&=& \big(  \gsp_3 \op V^{14}(\pi_2) \op V^{14}(\pi_2) \,,\, \l=-1 \big)\  
 \op\ \big(V^{14}(\pi_2)\,,\, \l= 3\big)
\nonumber \\
&&   \op\ ( \bC \o\,,\, \l= 5 ) \
   \op\  (\bR \o_0\,,\, \l=-3 )\ .
\label{La2sp3}\end{eqnarray}
Here, $\o$ is the symplectic form \re{sympl} and $\o_0$ the metric form 
$\o_0 = g_{\a\bb} dz^\a \wedge dz^{\bb}$.    $V^{14}(\pi_2) $ denotes the 
14-dimensional representation with highest weight $\pi_2$, the 2nd fundamental 
weight of $\gsp_3$.  
The 4-form $T_{(12)} $ is in fact one of six Sp(3)-invariant 4-forms in 12 dimensions.
The 17 equations \re{re12a}-\re{re12d} project the curvature two-form to the
49-dimensional eigenspace with eigenvalue $\l=-1$. The other eigenspaces 
have rather small dimensions compared with  $\dim(\La^2\bR^{12}) = 66$. 
We therefore expect the corresponding solutions to be rather trivial.
Sp(3), the stabiliser of the 4-form $T_{(12)}$ is a maximal subgroup of
SU(6).

The similarity of the equations \re{re12}-\re{hol12} to the three and six dimensional
systems  in $\bR^3$ and $\bC^3\simeq \bR^6$ discussed above  suggests that this 
is the counterpart  in  three dimensional quaternionic space $\bH^3 \simeq \bC^6$.
The imaginary quaternion units satisfy $i^2=j^2=k^2=-1$ and $ij=-ji=k$, together 
with the relations which result on cyclically permuting $(i,j,k)$. We consider 
$\bC$ to be an $\bR$-vector space spanned by $(1,i)$ and 
$\bH$ a $\bC$-vector space spanned by $(1,j)$.  Scalar multiplication of
$z\in\bC$ with the quaternionic basis element $j$ satisfies $zj = j \bz$, so quaternions
may be written in the form  
\begin{equation}
q :=   z + j \ol{w}  = z + w j \,,\quad  q\in \bH \,,\  z,w\in \bC\,.
\label{q}\end{equation}
The conjugate quaternion is then given by
\begin{equation}
\ol q :=   \bz -  {w}j  = \bz - j\ol w  \,,\quad  q\in \bH \,,\  z,w\in \bC\ .
\label{qconj}\end{equation}
The conjugate imaginary units are clearly given by 
$\ \ol i = -i \,,\  \ol  j = -j \,,\  \ol k = -k \,$.
Quaternions being noncommutative,  conjugation is an involutive antiautomorphism, 
i.e. $\ol{\ol q} = q$  and $\ol{q_1 q_2} = \ol q_2 \ol q_1$.
There exist related involutive automorphisms given by conjugation with the quaternion
units,
\begin{eqnarray}
\id :\ q   &\mapsto&\quad\  q= z + w j  \,, \nonumber\\  
 \a:\ q  &\mapsto&  -iqi  = z - w j   \,,\nonumber\\    
 \b:\ q   &\mapsto&  -jqj  = \bz + \ol w j \,,\nonumber\\   
 \g:\ q  &\mapsto&  -kqk  = \bz - \ol w j \,,   
\label{abc}\end{eqnarray}
in terms of which the real and imaginary parts of q can be expressed as linear
combinations of $q,\a(q),\b(q),\g(q)$ (see e.g. \cite{s}).

Now, let $M$ be a three quaternionic-dimensional (i.e. 12 real-dimensional) space.
In a local coordinate frame $T_pM \simeq \bH^3 \simeq  \bC^6$.
We define three quaternionic coordinates $q^A \,,\ A=1,2,3$, in terms of pairs of 
the complex coordinates 
$z^\a :=  x^\a + i x^{\a+6}\,,\, \a=1,\dots,6 $  used above,
\begin{eqnarray}
q^1 &:=&  z^1 + z^2 j \ =\  x^1 + i x^7 +j x^2 + k x^8\,,  \nonumber \\
q^2 &:=&  z^3 + z^4 j\ =\  x^3 + i x^9 +j x^4 + k x^0\,,  \nonumber \\
q^3 &:=&  z^5 + z^6 j \ =\  x^5 + i x^a +j x^6 + k x^b
\label{q12}\end{eqnarray}
and we denote the conjugate coordinates as $\ol{q^A} = q^{\ol A}$.

For any two quaternionic vector fields $Q_1, Q_2$ 
the curvature components $F(Q_1,  \g(Q_2))$ and   
$F(Q_2, \g(Q_1))$ have the same content in terms of real  curvature components,
since $\g$ is an involutive automorphism.
We now denote the basis vectors of the coordinate vector fields on $M$ by 
$Q_A := \p/\p q^A$, their quaternionic conjugates by 
$Q_{\ol A} := \ol{Q_A} = \p/\p q^{\ol A}$ and their $\a,\b,\g$-conjugates by
$Q_{\a(A)} := \a(Q_A)$, etc.
The hermitian metric in local quaternionic coordinates is given by 
$d^2s = g_{A\ol B} dq^A dq^{ \ol B}
 = dq^1 d  q^{\ol 1} + dq^2 d  q^{\ol 2} + dq^3 d  q^{\ol 3} $.

\bp
On a three quaternionic dimensional Riemannian manifold, the following
8 quaternionic curvature constraints are equivalent to 
the system \re{re12a}-\re{re12c} of self-duality equations in 12 dimensions:
\begin{eqnarray}
g^{A\ol B}\   F(Q_{\ol B}\,,\, Q_{\a(A)}) 
\ =\  \sum_{A=1}^3   F(Q_{\ol A}\,,\, Q_{\a(A)}) &=&   0   
\la{req12a} \\[4pt]
g^{A\ol B}\   F(Q_{\ol B}\,,\, Q_{\b(A)})  
\ =\  \sum_{A=1}^3   F(Q_{\ol A}\,,\, Q_{\b(A)})&=&   0   
\la{req12b} \\[8pt]
F(Q_{\ol 1}\,,\, Q_{\g(2)})\ =\  F( Q_{\ol 2}\,,\, Q_{\g(3)}) 
\ =\  F(Q_{\ol 3}\,,\, Q_{\g(1)}) &=& 0 
\la{holq12} \\[8pt]
F(Q_{\ol 1}\,,\, Q_{\g(1)}) \ =\   F( Q_{\ol 2}\,,\, Q_{\g(2)}) 
\ =\  F(Q_{\ol 3}\,,\, Q_{\g(3)}) &=& 0\,.  
\label{req12c} 
\end{eqnarray}
\ep 

\pf
The equivalence to the 17  equations  \re{re12a}-\re{re12c}, or equivalently to
the complex form \re{re12}-\re{hol12}  follows from a direct expansion of the
quaternionic vector fields in the basis $(1,j)$. \qed

\section{Flowing to 14 dimensions}

The similarity between the 3 quaternionic equations in \re{holq12}, 
the 3 complex equations in \re{hol6} and the 3 real equations in \re{d3}
immediately suggests that in analogy to the  flows \re{hf3}  and \re{hf7}, 
we may write down flows
for the three quaternionic partial curvatures in  \re{holq12} into a further complex 
direction, with coordinate $z^7$. We write,  in $M= M^3_{\bH} \times \bC$
with coordinates $(q^1,q^2,q^3, z^7)$, in analogy with \re{hf8cx} and \re{hf8re},
\begin{eqnarray}
\der{z^7} A(Q_1)  &=& - F( Q_{\ol 2}\,,\, Q_{\g(3)})  
\nonumber\\
\der{z^7}  A(Q_2)  &=& - F( Q_{\ol 3}\,,\, Q_{\g(1)})  
 \nonumber\\
\der{z^7}  A(Q_3)  &=& - F( Q_{\ol 1}\,,\, Q_{\g(2)}) 
 \nonumber\\
\der{z^7}  A(Z_{\ol 7})  &=&   \sum_{A=1}^3   F(Q_{\ol A}\,,\, Q_{\a(A)})
\la{hf14}\end{eqnarray}
together with  \re{req12b}, considered as an equation in 14 dimensions,
\begin{equation}
 \sum_{A=1}^3   F(Q_{\ol A}\,,\, Q_{\b(A)})  =   0   
\la{re14b} \,.  
\end{equation}
Writing the quaternionic vector fields $Q_A\,,\  A=1,\dots,3$ in terms of complex 
vector fields  $Z_\a\,,\   \a=1,\dots,6$ 
according to the choice in  \re{q12}  and unravelling the $A(Z_7)=0$ gauge, 
we obtain the system
\begin{eqnarray}
 F(Z_{7} \,,\,Z_{1}+Z_{2}j)  \ +\   F(\ol{Z_3} - j\ol{Z_4} \,,\,\ol{Z_5}- \ol{Z_6}j)  &=&  0  
\nonumber \\
  F(Z_{7} \,,\,Z_{3}+Z_{4}j)   \ +\   F(\ol{Z_5} - j\ol{Z_6} \,,\,\ol{Z_1}- \ol{Z_2}j) &=&  0    
\nonumber \\
F(Z_{7} \,,\,Z_{5}+Z_{6}j)   \ +\   F(\ol{Z_1} - j\ol{Z_2} \,,\,\ol{Z_3}- \ol{Z_4}j) &=&  0  
   \nonumber\\
F(Z_{7} \,,\, \ol{Z_{7}})  \ -\   
 \sum_{\a=1}^3   F(\ol{Z_{2\a -1}} - j\ol{Z_{2\a}} \,,\, Z_{2\a -1} - j\ol{Z_{2\a}})&=&  0  
 \nonumber\\
 \sum_{\a=1}^3   F(\ol{Z_{2\a -1}} - j\ol{Z_{2\a}} \,,\, \ol{Z_{2\a -1}} + j Z_{2\a})&=&  0
\la{d14q} 
\end{eqnarray}
Expanding the quaternionic vector fields in the basis $(1,j)$, we obtain  
equations on $\bC^7$, which are contained in the system
\begin{eqnarray}
 F_{\a\b} +  \half   \F_{\a\b\g\d}\  F_{\bg\bd}  = 0
\label{hol14}\\
g^{\a\bb} F_{\a\bb}  = 0
\label{re14}
\end{eqnarray}
with   $\F$ given by the G$_2^\bC$-invariant  (4,0)-form
\begin{equation}
\F= dz^{1234} + dz^{1256} +dz^{3456} +dz^{1375} +dz^{1467} +dz^{2367} +dz^{2457}.  
\label{F14}\end{equation}
By Theorem \ref{cx_thm} we therefore have a system of equations which
implies the Yang-Mills equations  in 14 dimensions.

Unlike the previous analogous cases, the equations \re{hol14} are not equivalent
to the set \re{d14q}. The former set contains more equations than the latter.
More precisely, \re{hol14} includes, for instance, the three equations
\begin{equation}
 F_{71} + F_{\bar3\bar5}+F_{\bar6\bar4} \ =\ F_{71} + F_{\bar3\bar5}+F_{64} 
 \ = \  F_{71} + F_{35}+F_{\bar6\bar4} = 0\,.
\label{ex}\end{equation} 
Under the G$_2^\bC$-invariant 4-form  $\F$, both real and imaginary parts of 
$F_{\a\b}$ split into their 7- and 14-dimensional irreducible parts. The  
equations of the form \re{ex} imply that under \re{hol14} the real part is 
projected to the 14-dimensional piece (7 equations) and the imaginary part is
zero (21 equations). The real form of the system \re{hol14},\re{re14} is given 
by the set of 29 equations, 
\begin{eqnarray}
 F_{18}+F_{29}+F_{30}+F_{4a}+F_{5b}+F_{6c} +F_{7d} &=& 0                       
\nonumber\\                         
 F_{12}+F_{34}+F_{56}-F_{89}-F_{0a} -F_{bc} &=& 0
\nonumber 
\\
 F_{13}-F_{24}-F_{80}+F_{9a}+F_{bd}  +F_{75} &=& 0                        
\nonumber\\  
 F_{14}+F_{23}-F_{8a}-F_{90}-F_{cd}-F_{76}   &=& 0      
\nonumber\\  
F_{15}-F_{26}-F_{8b}+F_{9c}-F_{0d}-F_{73}    &=& 0            
\nonumber\\ 
 F_{16}+F_{25}-F_{8c}-F_{9b}+F_{ad} +F_{74}  &=& 0     
\nonumber\\  
F_{17} - F_{35}+F_{46}-F_{8d}+F_{0b}-F_{ac}  &=& 0       
\nonumber\\
F_{72}  -F_{36}-F_{45}+F_{9d}+F_{0c}+F_{ab}  &=& 0                                         
\la{d14_6}\\[4pt]  
F_{78} - F_{1d} \,\ =\,\                         
F_{79} - F_{2d}  \,\ =\,\                           
F_{70} - F_{3d}  &=& 0                      
\nonumber\\  
F_{7a} -  F_{4d}   \,\ =\,\                          
 F_{7b} -  F_{5d}  \,\ =\,\                        
F_{7c} -  F_{6d}   &=& 0                     
\nonumber\\  
 F_{19}-F_{28}   \,\ =\,\                          
 F_{10} -F_{38}   \,\ =\,\                          
F_{1a}-F_{48}      &=& 0                  
\nonumber\\  
 F_{1b}-F_{58}     \,\ =\,\                       
 F_{20}-F_{39}        \,\ =\,\                   
F_{2a} -F_{49}     &=& 0                   
\nonumber\\  
 F_{3a}- F_{40}    \,\ =\,\                        
 F_{2b}-F_{59}    \,\ =\,\                        
F_{3b}-F_{50}      &=& 0                  
\nonumber\\  
F_{4b}-F_{5a}        \,\ =\,\                    
F_{1c} - F_{68}        \,\ =\,\                    
F_{2c} - F_{69}      &=& 0                  
\nonumber\\ 
F_{3c} - F_{60}       \,\ =\,\                     
 F_{4c}- F_{6a}         \,\ =\,\                   
 F_{5c}-F_{6b}         &=& 0 \,.              
\la{d14_21} 
\end{eqnarray}
These 29 equations correspond to the $\l =-1$ eigenspace of the special democratic 
4-form   given by
\begin{eqnarray*}
T_{(14)} &=&\quad dx^{1234} +   dx^{1256} +   dx^{1298} +   dx^{12a0} +   dx^{12cb} + 
dx^{1375} +   dx^{13bd}  
\nonumber\\  &&+\   dx^{1308} +   dx^{139a} +  dx^{14a8} +   
dx^{1409} +   dx^{1467} +    dx^{14dc} +  dx^{15b8}  
\nonumber\\  &&+\    dx^{159c} +    
dx^{15d0} +   dx^{16c8} +  dx^{16b9} +    dx^{1ad6} +   dx^{170b} +  
dx^{1a7c}  
\nonumber\\  &&+\     dx^{2367} +   dx^{23a8} +  dx^{2309} +     dx^{23dc} + 
 dx^{2457} +   dx^{2480} +   dx^{24a9}  
\nonumber\\  &&+\     dx^{24db} +   dx^{25c8} +   
 dx^{25b9} +  dx^{2ad5} +  dx^{268b} +   dx^{26c9} +   dx^{20d6}  
\nonumber\\  &&+\    
 dx^{207c} +  dx^{2a7b} +    dx^{3456} +    dx^{3498} +   dx^{34a0} +  
 dx^{34cb} +  dx^{35b0}  
\nonumber\\  &&+\     dx^{35ac} +  dx^{835d} +   dx^{93d6} +    
 dx^{36c0} +  dx^{36ba} +   dx^{83b7} +   dx^{937c}  
\nonumber\\  &&+\  dx^{94d5} +  
 dx^{45c0} +  dx^{45ba} +  dx^{460b} +   dx^{46ca} +   dx^{84d6} +   
dx^{847c}  
\nonumber\\  &&+\    dx^{947b} +  dx^{5698} +   dx^{56a0} +   dx^{56cb} +  
dx^{9a75} +   dx^{8a76} +   dx^{9076}  
\nonumber\\  &&+\   dx^{890a} +   dx^{89bc} +   
 dx^{8057} +    dx^{80db} +    dx^{8acd} +   dx^{90cd} +     dx^{9abd}  
\nonumber\\  &&+\      
 dx^{0abc} +  dx^{187d} +   dx^{297d} +   dx^{307d} +   dx^{4a7d} +   
 dx^{5b7d} +   dx^{6c7d}.    
\la{tt14}\end{eqnarray*}
Its characteristic polynomial is given by
\begin{equation}
\chi(T_{(14)}) =  (\l+1)^{62}   (\l-3)^{14}   (\l+ 3)^{7}   (\l-5)^{7}   (\l-6)
\label{char14}\end{equation}
and the above 29 equations correspond to the projection to 62-dimensional
$\l {=}-1$ eigenspace.

Deleting all terms containing  the 14th index $d$ from the above equations
yields the 13-dimensional reduction, corresponding to a flow along a real
parameter rather than the complex one chosen in  \re{hf14}. This is
 also a set of 29 equations, projecting 
the curvature to the 49-dimensional $\l=-1$ eigenspace of the corresponding 
reduction of the  4-form $T_{(14)}$. The reduced 4-form has  
characteristic polynomial
\begin{equation}
\chi(T_{(13)}) =  
(\l+1)^{49}   (\l-3)^{8}   (\l + 3) (\l-5)^2 (\lambda-4)^6 (\lambda^2+\lambda-4)^6\,.
\label{char13}\end{equation}

\section{Oxidation to 16 dimensions} 
Analogously to the oxidations \re{hf8}, \re{hf8cx} and \re{hf8re} 
to eight real dimensions, we may oxidise  the system \re{hol14},\re{re14} in $\bC^7$ 
to one in $\bC^8$ by taking $g_{\a\bb}$ to be the $\bC^8$-metric 
and the  (4,0)-form $\F$ to be given by the Spin(7)$^\bC$-invariant,
\begin{eqnarray}
\F &=&\ dz^{1234} + dz^{1256} + dz^{1278} +dz^{3456} + dz^{3478}  + dz^{5678} +dz^{1368}\\
   && +dz^{1375} +dz^{1467} +dz^{1458} +dz^{2367} +dz^{2457}+dz^{2358} +dz^{2486}.  
\label{F16}\end{eqnarray}
The corresponding system includes the flow equations based on  \re{d14q},
 \begin{eqnarray}
 F(Z_{8} \,,\,Z_{1}j-Z_{2})  \ +\   F(Z_{7} \,,\,Z_{1}+Z_{2}j)  \ +\   F(\ol{Z_3} - j\ol{Z_4} \,,\,\ol{Z_5}- \ol{Z_6}j)  &=&  0  \nonumber \\
 F(Z_{8} \,,\,Z_{3}j-Z_{4})  \ +\    F(Z_{7} \,,\,Z_{3}+Z_{4}j)   \ +\   F(\ol{Z_5} - j\ol{Z_6} \,,\,\ol{Z_1}- \ol{Z_2}j) &=&  0    
\nonumber \\
 F(Z_{8} \,,\,Z_{5}j-Z_{6})  \ +\  F(Z_{7} \,,\,Z_{5}+Z_{6}j)   \ +\   F(\ol{Z_1} - j\ol{Z_2} \,,\,\ol{Z_3}- \ol{Z_4}j) &=&  0  
   \nonumber\\
F(Z_{8} \,,\, \ol{Z_{8}})  \ +\   F(Z_{7} \,,\, \ol{Z_{7}})  \ -\   
 \sum_{\a=1}^3   F(\ol{Z_{2\a -1}} - j\ol{Z_{2\a}} \,,\, Z_{2\a -1} - j\ol{Z_{2\a}})&=&  0  
 \nonumber\\
 \sum_{\a=1}^4   F(\ol{Z_{2\a -1}} - j\ol{Z_{2\a}} \,,\, \ol{Z_{2\a -1}} + j Z_{2\a})&=&  0\,.
\la{d16q} 
\end{eqnarray}

The real form of the full system of equations with (4,0)-form $\F$ given in \re{F16}
is given by
\begin{eqnarray}
F_{12} +  F_{34} +  F_{56} +  F_{78}- F_{90}- F_{ab}- F_{cd}- F_{ef} &=& 0
\nonumber\\
F_{13}- F_{24}- F_{57} +  F_{68}- F_{9a} +  F_{0b} +  F_{ce}- F_{df} &=& 0
\nonumber\\
F_{14} +  F_{23} +  F_{58} +  F_{67}- F_{9b}- F_{0a}- F_{cf}- F_{de} &=& 0
\nonumber\\
F_{15}- F_{26} +  F_{37}- F_{48}- F_{9c} +  F_{0d}- F_{ae} +  F_{bf} &=& 0
\nonumber\\
F_{16} +  F_{25}- F_{38}- F_{47}- F_{9d}- F_{0c} +  F_{af} +  F_{be} &=& 0
\nonumber\\
F_{17}- F_{28}- F_{35} +  F_{46}- F_{9e} +  F_{0f} +  F_{ac}- F_{bd} &=& 0
\nonumber\\
F_{18} +  F_{27} +  F_{36} +  F_{45}- F_{9f}- F_{0e}- F_{ad}- F_{bc} &=& 0
\nonumber\\
F_{19} +  F_{20} +  F_{3a} +  F_{4b} +  F_{5c} +  F_{6d} +  F_{7e} +  F_{8f}&=& 0 
\nonumber\\[4pt]
F_{79} - F_{1e}  \ =\       F_{80} - F_{2f}  
     \ =\   F_{5a}  - F_{3c}  \ =\  F_{6b} -F_{4d}   &=& 0 
\nonumber\\
F_{70}  - F_{2e}     \ =\  F_{3d} -F_{6a} \ =\   
  F_{4c}-F_{5b} \ =\  F_{1f} -F_{89} &=& 0 
\nonumber\\
F_{69} -F_{1d}  \ =\   F_{2c}-F_{50}
 \ =\  F_{3f} -F_{8a} \ =\    F_{7b}-F_{4e}   &=& 0  
\nonumber\\
F_{1c}-F_{59}  \ =\  F_{2d} - F_{60}  
 \ =\  F_{3e}-F_{7a}   \ =\   F_{8b} - F_{4f} &=& 0
\nonumber\\
F_{1b} -F_{49}  \ =\  F_{5f}-F_{8c} \ =\  
F_{2a} - F_{30}  \ =\   F_{6e}-F_{7d}&=& 0
\nonumber\\
F_{6f}-F_{8d} \ =\  F_{5e}-F_{7c} 
 \ =\  F_{40} - F_{2b} \ =\  F_{39}-F_{1a} &=& 0
\nonumber\\
 F_{7f}-F_{8e} \ =\   F_{5d}-F_{6c}  
 \ =\  F_{10}-F_{29} \ =\   F_{3b}-F_{4a} &=& 0\,. 
\end{eqnarray}
The corresponding 4-form  $T_{(16)}\in \Lambda^4\bR^{16}$ has characteristic polynomial
\begin{equation}
\chi(T_{(16)}) =  (\l+1)^{84}   (\l-3)^{21}  (\l-7)^{8}   (\l+5)^{7} \,,  
\label{char16}\end{equation}
so the above 36 equations correspond to the vanishing of the imaginary part of 
$F_{\a\b}$ (28 equations), the 7-dimensional irreducible piece of the real part of 
$F_{\a\b}$ and the singlet trace condition on the (1,1)-curvature.

Deleting all terms containing $f$, the 16th index, from the above equations
yields 36 equations in 15 dimensions which projects the curvature to the 
69-dimensional $\l=-1$ eigenspace of the corresponding 4-form $T_{(15)}$,
which has characteristic polynomial
\begin{equation}
\chi(T_{(15)}) = 
(\lambda+1)^{69} (\lambda-6)^8 (\lambda-3)^{14} (\lambda^2+3\lambda-6)^7\,. 
\label{char15}\end{equation}

\section{The reductions to $8 < d < 12$}
We  now briefly comment on some reductions of
the above 12-dimensional system to the lower dimensions which were 
missed out in the discussion above.
\bigskip

\noindent
{\bf d=11}
\medskip

\noindent
Deleting all terms containing $dx^b$ in \re{re12a}-\re{re12d} yields a set of
17 equations in 11-dimensions. The correspondingly reduced four-form 
$T_{(11)}:=T_{(12)}\vert_{dx^b=0}$  has  characteristic polynomial
\begin{equation}
\chi(T_{(11)}) = (\l +1)^{38} (\l -2)^{8} (\l -3)^{5} (\l -4)^2 (\l^2+\l-4) .
\label{char11}\end{equation}

\noindent
{\bf d=10}
\medskip

\noindent 
Reducing the above
11-dimensional 4-form further to the 10-dimensional hypersurface defined,
for instance, by $x^6 = 0$ yields a 4-form with  characteristic polynomial
\begin{equation}
\chi(T_{(10)}) = (\l +1)^{30} (\l -1)^{8} (\l -3)^{6} (\l -4)  .
\label{char10}\end{equation}
The $\l=-1$ eigenspace corresponds to a set of 15 equations amongst the 
45 curvature components.
This case is the complex counterpart of the $d=5$ case discussed at the
end of section \ref{sect_d6}. 
In $\bC^5$, these equations take the form \re{hol},\re{re} with $\a,\b = 1,\dots,5$
and the complex (4,0)-form given by the contraction of the (5,0) volume form with 
a constant unit (0,1)-vector.  This (4,0)-form is the SU(4)-invariant volume form 
in the 4-dimensional complex space orthogonal to this vector. Choosing, this vector, 
for instance in the direction of the $z^5$-axis, we obtain $\F = dz^{1234}$, 
yielding the following equations on $\bC^5$
\begin{eqnarray}
&& F_{1\bar1} + F_{2\bar2} + F_{3\bar3} + F_{4\bar4} + F_{5\bar5}  = 0 \nonumber\\[5pt]
&& F_{12} + F_{\bar3 \bar4} =  F_{13} + F_{\bar4 \bar2} =  F_{14} + F_{\bar2 \bar3} = 0  \nonumber\\[5pt]
&&F_{15} =  F_{2 5} =  F_{35}  = F_{45} = 0 \,.
\label{d10a}
\end{eqnarray}

\bigskip

\noindent
{\bf d=9}
\medskip

\noindent
 
The most symmetric reduction of  \re{d10a} to 9-dimensions, making $z^5$ real,
is a trivial embedding of the Spin(7)-invariant set of equations  \re{d8c}
in 9 dimensions.

\section{Some open questions}

An intriguing open problem is the relation of the 12-dimensional
system to sextonions and to the `missing row' of the Freudenthal magic square
related to $E_{7\half}$ (see \cite{w,lm}).

It remains to be seen whether interesting solutions to the higher dimensional
equations presented here can be found. The simplest solutions are embeddings 
of known solutions in dimensions $d \le 8$ in the higher dimensional theories.
Do such solutions allow oxidation to nontrivial solutions in higher dimensions?
In particular, when the duality equations describe (partial) curvature flows, 
to what  extent do solutions in the bulk  arise from solutions on the 
initial value surface (boundary) of the flow?
For instance,  can the known four-dimensional solutions of the 
self-duality equations \re{d4} be seen as arising from a flow which has a flat 
3d connection as its initial value, or do the known solutions of the 
8-dimensional Spin(7)-invariant equation \cite{fn,funi,hilp}
arise as  solutions of the flow equation  \re{hf8} from solutions (e.g. \cite{gn}) 
of the G$_2$-invariant equation \re{d7} on the initial value seven-fold?

\section*{Acknowledgements}
This work has benefitted a great deal from innumerable discussions over the
last 30 years with Jean Nuyts. 
I acknowledge useful discussions with Andrea Spiro and Gregor Weingart,
as well as  partial funding from the SFB 647 ``Raum-Zeit-Materie''
of the Deutsche Forschungsgemeinschaft. 
I should like to thank Hermann Nicolai and the Albert-Einstein-Institut for hospitality.

\end{document}